\documentclass[aps,pra,twocolumn]{revtex4}

\usepackage{amsmath}

\usepackage{amsfonts}

\usepackage{amssymb}

\usepackage{graphicx}

\usepackage{color}
 \usepackage{bbm} 
\usepackage{appendix}
\providecommand{\U}[1]{\protect\rule{.1in}{.1in}}

\newtheorem{theorem}{Theorem}

\newtheorem{criterion}[theorem]{Criterion}

\newenvironment{proof}[1][Proof]{\noindent\textbf{#1.} }{\ \rule{0.5em}{0.5em}}

\begin{document}

\title{An alternative representation for  pure symmetric states of qubits and its applications to entanglement classification}

\author{A. Mandilara$^1$$^2$, T. Coudreau$^1$, A. Keller$^3$ and P. Milman$^1$}

\affiliation{$^{1}$Laboratoire Mat\'eriaux et Ph\'enom\`enes Quantiques, Sorbonne Paris Cit\'e, Universit\'e Paris Diderot, CNRS UMR 7162, 75013, Paris, France}
\affiliation{$^{2}$Nazarbayev University, Department of Physics, School of Science and Technology, 53 Kabanbay Batyr Avenue, 010000, Astana, Republic of Kazakhstan}
\affiliation{$^{3}$Univ. Paris-Sud 11, Institut de Sciences Mol\'eculaires d'Orsay (CNRS), B\^{a}timent 350--Campus d'Orsay, 91405 Orsay Cedex, France}

\begin{abstract}

We prove that  the vast majority of symmetric states of qubits (or spin $1/2$) can be decomposed in a unique way into a superposition of spin $1/2$ coherent states. For  the case of two qubits, the proposed decomposition  reproduces the Schmidt decomposition and therefore, in the case of a higher number of qubits,  can be considered as its generalization. We analyze the geometrical aspects of the   proposed representation and its invariant properties  under the action of local unitary and local invertible transformations. As an application, we  identify the most general classes of entanglement and representative states for any number of qubits in a symmetric state. 
\end{abstract}

\maketitle

 Symmetric states under permutations have drawn lately a lot of attention in the field of quantum information. The essential reason is that the number of parameters needed for the description of a state in a symmetric subspace scales just linearly with the number of parties. This simplification makes symmetric states a good test-ground for complex quantum information tasks such as the description of multipartite entanglement \cite{EntSG, Aulbach, Solano1, Solano2, Martin,Damian, Damian2, Lyons} and quantum tomography \cite{Tomography}.

There are two well known  representations for symmetric states, the Dicke basis \cite{Dicke} and Majorana representation \cite{Majorana}.  In the present work, we introduce  a novel representation  whose forms resembles strongly to a generalized Schmidt decomposition and that presents advantages with respect to the previous ones  as regards the readability of the entanglement properties of the state. The structure of the proposed representation remains invariant under the action of local unitary operations which leave the state in the symmetric subspace.  We associate to the proposed decomposition a geometric representation of the states that  has the interest of displaying this  invariance. We proceed by identifying invariant forms of the representation under the action of local unitary and local invertible transformations. In this way we arrive to an exact classification of entanglement for symmetric states. With the well studied example of three qubits we establish a first connection among the suggested representation and measures of multipartite entanglement. Furthermore, an immediate consequence of  our methods is  a straightforward estimation of a well established measure of entanglement, the so called Schmidt measure \cite{SN} for the symmetric states.

The structure of this paper is the following. We start by recalling some basic properties of the Dicke and Majorana representation and their connection to entanglement classification. We then move to the presentation of the novel decomposition that is the main result of the paper. Finally we employ the properties of the decomposition in order to arrive at a classification of entanglement applicable to the vast majority of symmetric states.

Every symmetric state of $N$ qubits can be expressed in a unique way over the Dicke basis formed by the $N+1$ joined eigenstates $\left\{  \left\vert N/2,m\right\rangle \right\}  $ of the collective operators $\hat{S}_{Z}={\textstyle\sum\limits_{i=1}^{N}}\hat{\sigma}_{i}^{Z}$ and 
$\hat{S}^2$, where $\vec{\hat{S}} =  {\textstyle\sum\limits_{i=1}^{N}}\vec{\hat{\sigma}}_{i}$:

\begin{equation} \left\vert N/2,m\right\rangle =d_{N,m}^{-1}\sum_{perm}\underset{m+N/2}{\underbrace{\left\vert 1\right\rangle \left\vert 1\right\rangle \ldots\left\vert 1\right\rangle }}\underset{N/2-m}{\underbrace{\left\vert 0\right\rangle \left\vert 0\right\rangle \ldots\left\vert 0\right\rangle }}\end{equation} where $d_{N,m}=\tiny{\sqrt{\text{$\left( \begin{array} [c]{c} N\\ m+N/2 \end{array} \right)  $}}}$. More specifically $\hat{S}_{Z}\left\vert N/2,m\right\rangle=m\left\vert N/2,m\right\rangle $ $\ $and $\hat{S}^{2}\left\vert N/2,m\right\rangle =N\left(  N/2+1\right)  /2\left\vert N/2,m\right\rangle $ with $m=-N/2,-N/2+1,...,N/2$. Even if the number of $\hat{S}_{Z}$ eigenstates is $N+1$, using the freedom of choice of the global phase and the normalization condition, one remains with $N$ complex numbers expressing in a unique way every symmetric state over this basis.

A commonly used alternative  to the Dicke basis is the Majorana representation \cite{Majorana} initially proposed to describe states of spin-$j$ systems. This attributes to each state $N$ points on the Bloch  sphere in the following way: one projects the given symmetric state $\left\vert \Psi\right\rangle =$ $\sum_{m=-N/2}^{N/2}c_{m}\left\vert N/2,m\right\rangle $ on a spin coherent state \cite{Radcliffe} of $N$ qubits defined as \begin{align} \left\vert \alpha\right\rangle  &  =e^{\alpha^{\ast}\hat{S}_{+}}\left\vert N/2,-N/2\right\rangle \label{coh1}\\ &  =\sum_{m=-s}^{s}\left(  \alpha^{\ast}\right)  ^{N/2+m}d_{N,m}\left\vert N/2,m\right\rangle . \end{align}

This projection leads to a polynomial of $Nth$ order on the complex parameter $\alpha$, the so-called Majorana polynomial:\begin{equation} \Psi\left(  \alpha\right)  =\left\langle \alpha\right.  \left\vert \Psi\right\rangle =\sum_{m=-N/2}^{N/2}\lambda_{m}\alpha^{N/2+m} ,\label{MajoPoly}  \end{equation}
with $\lambda_{m}=d_{N,m}c_{m}$. The $N$ complex roots $\left\{  \alpha_{n}\right\}  $ (Majorana roots) of the polynomial $\Psi\left(\alpha\right)$ \begin{equation} \Psi\left(  \alpha\right)  \propto{\textstyle\prod\limits_{n=1}^{N}}\left(  \alpha-\alpha_{n}\right)  .\label{MajoRoots} \end{equation}
fully characterize the state $\left\vert \Psi\right\rangle $. It is possible to introduce a geometric picture by attributing  \ $N$ Bloch vectors $\left\{ \mathbf{v}_{n}\right\}  $ to the roots $\left\{  \alpha_{n}\right\}  $ via the inverse stereographic mapping $\left\{  \alpha_{n}\right\}  \rightarrow \left\{  e^{i\varphi_{n}}\tan\left(  \theta_{n}/2\right)  \right\} \equiv\left\{  \mathbf{v}_{n}\right\}  $. The edges of these vectors define a set of $N$ points on the Bloch sphere, the so called Majorana stars. Furthermore, using the states $\left\vert \chi_{n}\right\rangle =\sin\left(  \theta_{n}/2\right)  \left\vert 0\right\rangle -e^{-i\varphi_{n}}\cos\left(  \theta_{n}/2\right)  \left\vert 1\right\rangle $ which are orthogonal to the states  associated to $\left\{  \mathbf{v}_{n}\right\} $, one can write in a unique way every symmetric state as \begin{equation} \left\vert \Psi\right\rangle =A\sum_{perm}\left\vert \chi_{1}\right\rangle \otimes\left\vert \chi_{2}\right\rangle \otimes\ldots\left\vert \chi_{N}\right\rangle \label{Majo2} \end{equation} where $A$  is  a complex number that stands for the normalization factor and the global phase.

Majorana representation,  among other applications \cite{Kolederski, Markela, Bruno, Wang}, has  proven very useful  to the study and classification of  entanglement in symmetric states \cite{Aulbach, Solano1,Solano2,Martin,Damian,Damian2, Mosseri}. There are two ways to classify entanglement or in other words, to regroup states in  classes according to their entanglement properties. In the first classification,  each class contains states which can be transformed in each others  by  Local Unitary ($LU$) transformations. States belonging to the same, so-called, $LU$ class of entanglement have identical entanglement properties.  In the second classification, Stochastic Local Operations and Classical Communication (\textit{SLOCC}) are also allowed. In that case, it can be shown \cite{Dur,Verstraete} that two states are in the same class if and only if one state can be converted to the other via the use of  Invertible Local ($IL$) operations  mathematically implemented by the $SL(2,\mathbb{C})$ group. States belonging to same so-called $IL$ class of entanglement, are entangled in the same way.

Focusing on the case of symmetric states of qubits the $LU$ transformations which  leave states in the symmetric subspace are equivalent to collective $\mathbf{SU(2)}$ rotation ($\mathbf{SU(2)}=SU(2)\times SU(2)\ldots\times SU(2)$) \cite{Solano2}  where all the $SU(2)$ (single qubit) transformations are identical and are parametrized by $3$ real numbers. A symmetric state of qubits is defined by $2N$ real parameters  but identifying its  invariant part under $LU$ \ transformations requires only $2N-3$ real numbers, the so-called $LU$ invariants.  States with the same $LU$ invariants  belong to the same $LU$  entanglement class. There are different ways of identifying a set of $LU$ invariants for a given state \cite{Gingrich}. One way is to calculate the values of a complete set of polynomial invariant quantities. Alternatively with the help of $LU$ transformations one can reduce a given state to a properly chosen $LU$ canonical form \cite{Mandilara}, described by $2N-3$ real numbers. For symmetric states, the Majorana representation offers an overcomplete set of $LU$ invariants with a geometric aspect, the inner products among $\left\{\mathbf{v}_{n}\right\}$. This can be easily understood noting an essential aspect of Majorana representation: Majorana stars rotate uniformly under $LU$ transformations (see Eq.(\ref{Majo2})).

In the case of $IL$ transformations for symmetric states, it has been proven~\cite{Solano2} that is sufficient to search for interconvertibility  via just collective $\mathbf{SL(2,\mathbb{C})}$ operations i.e. $\mathbf{SL(2,\mathbb{C})=}SL(2,\mathbb{C})\times SL(2,\mathbb{C})\ldots\times SL(2,\mathbb{C})$. The $SL(2,\mathbb{C})$ group is the complexification of the $SU(2)$ group and thus a collective $IL$ transformation on a symmetric state is parametrized by $6$ real numbers.  Similarly to the case of\ $LU$ transformations, the $2N-6$ invariants of a symmetric state under $IL$ transformations ($IL$ invariants), can be calculated in different ways. It has been recently proven that under $SL$ operations Majorana roots follows  M\"obius transformations on the complex plane \cite{Aulbach,Biju,Mosseri} and that a complete set of $IL$ invariants \cite{Mosseri} is given by combinations of the roots of the Majorana polynomials. Finally, it is worth mentioning that  a third way for classifying entanglement relevant only for symmetric states, has been recently suggested \cite{Solano1}. This last classification is related to the fact that an $IL$ transformation cannot change  the classification of degeneracies of the Majorana polynomial Eq.(\ref{MajoPoly}).   

In what follows we only consider pure states and we call \textit{generic} symmetric state of  $N$ qubits, a pure state whose highest degeneracy degree ($\gamma$) in Majorana's roots satisfies the condition $\gamma<\frac{N+1}{2}$ or $\gamma=N$   for $N$  odd, and $\gamma<\frac{N}{2}+1$ or $\gamma=N$ when $N$ is even. In the class of generic states, the states with no degeneracies are included which are the states covering the vast majority of space of symmetric states.   
\vspace{1cm}

We start by claiming that any generic symmetric state $\left\vert\Psi_{(odd)}\right\rangle$  of $N$ qubits where $N$ is \textit{odd} can be decomposed in a unique way as a superposition of at most $\left(N-1\right)/2$ spin coherent states $\left\vert \mathbf{\Phi}_{m}\right\rangle$: \begin{align}\left\vert\Psi_{(odd)}\right\rangle  &  =\left({\textstyle\sum\limits_{m=0}^{\left(  N-1\right)/2}}c_{m}\left\vert \mathbf{\Phi}_{m}\right\rangle \right)  \label{ini}\\ \left\vert \mathbf{\Phi}_{m}\right\rangle  &  =\left\vert \phi_{m}\right\rangle \otimes\left\vert \phi_{m}\right\rangle ...\left\vert \phi _{m}\right\rangle .\nonumber \end{align} As a convention we arrange the complex amplitudes $c_{m}$ in  decreasing sequence $\left\vert c_{0}\right\vert >\left\vert c_{1}\right\vert >...>\left\vert c_{\left(N-1\right) /2}\right\vert $ and for the single qubit states $\left\vert \phi_{m}\right\rangle$, we use the specific parametrization  $\left\vert \phi_{m}\right\rangle =\cos\left(\theta_{m}/2\right)  \left\vert 0\right\rangle +e^{i\varphi_{m}}\sin\left(\theta_{m}/2\right)  \left\vert 1\right\rangle $. We also note that in the general case $\left\langle \phi_{n}\right.  \left\vert \phi_{m}\right\rangle \neq0$.  \ If we exploit the normalization condition and the freedom of choice of the global phase, we can rewrite the state in the following more convenient form for our purposes, \begin{equation}\left\vert \Psi_{(odd)}\right\rangle =A\left(\left\vert \mathbf{\Phi}_{0}\right\rangle +{\textstyle\sum\limits_{m=1}^{\left(  N-1\right)/2}} y_{m}e^{ik_{m}}\left\vert \mathbf{\Phi}_{m}\right\rangle \right) \label{main}\end{equation} where $y_{m}=\left\vert c_{m}/c_{0}\right\vert <1$, $e^{ik_{m}}=c_{m}(c_{0}y_{m})^{-1}$ and $A$ the normalization factor. The proof of Eq.(\ref{ini}) which is presented in the Appendix,  additionally  provides the steps for identifying the parameters  of  decomposition (states $\left\vert \phi_{m}\right\rangle$ and coefficients $c_{m}$)   for a given  generic state.

For the case of an \textit{even} number of qubits, the decomposition is slightly different. We prove in the Appendix that the following unique decomposition exists  \begin{align} \left\vert \Psi_{(even)}\right\rangle  &  =c_{0}\left\vert \mathbf{\Phi}_{0}\right\rangle +c_{1}\left\vert \mathbf{\Phi}_{0}^{\bot}\right\rangle +{\textstyle\sum\limits_{\substack{m=2\\(N>2)}}^{N/2}}c_{m}\left\vert \mathbf{\Phi}_{m}\right\rangle \label{even}\\\left\vert \mathbf{\Phi}_{m}\right\rangle  &  =\left\vert \phi_{m}\right\rangle \otimes\left\vert \phi_{m}\right\rangle ...\left\vert \phi_{m}\right\rangle .\nonumber\end{align} 
The complex amplitudes $c_{m}$ satisfy now the following conditions: $\left\vert c_{2}\right\vert >...>\left\vert c_{N/2}\right\vert$ and $\left\vert c_{0}\right\vert >\left\vert c_{1}\right\vert $. In addition $\left\langle \phi_{0}\right.  \left\vert\phi_{0}^{\bot}\right\rangle =0$. For the single qubit states we use as before the convention $\left\vert \phi_{m}\right\rangle =\cos\left(\theta_{m}/2\right)\left\vert 0\right\rangle +e^{i\varphi_{m}}\sin\left( \theta_{m}/2\right)  \left\vert 1\right\rangle $. We suggest the more convenient  form \begin{equation} \left\vert \Psi_{(even)}\right\rangle =A\left(  \left\vert \mathbf{\Phi}_{0}\right\rangle +y_{1}e^{ik_{1}}\left\vert \mathbf{\Phi}_{0}^{\bot}\right\rangle +{\textstyle\sum\limits_{\substack{m=2\\(N>2)}}^{N/2}}y_{m}e^{ik_{m}}\left\vert \mathbf{\Phi}_{m}\right\rangle \right)\label{even2}\end{equation} where $y_{m}=\left\vert c_{m}/c_{0}\right\vert $, $e^{ik_{m}}=c_{m}(c_{0}y_{m})^{-1}$ and $A$ the normalization factor.

One interesting feature of the decomposition  Eq.(\ref{main})   is that it permits us to obtain a geometrical image for the state $\left\vert\Psi_{(odd)}\right\rangle$. We can represent Eq.(\ref{main}) on the Bloch  ball with one normalized vector $\left\vert \phi_{0}\right\rangle $ and $\left(  N-1\right)  /2$ unnormalized vectors $y_{m}\left\vert \phi_{m}\right\rangle $ of length ${\bf l}\leqslant1$. The only ingredient missing in the picture are the $\left(  N-1\right)  /2$ real phases $k_{m}$. Similarly, in the case of an even number of quits, one can attribute a  geometrical picture to the state in  Eq.(\ref{even2}), with normalized vector $\left\vert \phi_{0}\right\rangle $, the unnormalized vector $y_{1}\left\vert \phi_{0}^{\bot}\right\rangle $, and $N/2-1$ unnormalized vectors $y_{m}\left\vert \phi_{m}\right\rangle $.  This geometric picture remains invariant under the action of local  unitary transformations which leave the state permutationaly symmetric.

In order to show this, let us start with the case of an odd number of qubits and consider the state $\left\vert \Psi_{(odd)}\right\rangle $ in Eq.(\ref{main}). We denote  $\left\vert \Psi_{(odd)}^{\prime}\right\rangle =\hat{U}\left\vert \Psi_{(odd)}\right\rangle $  with $\hat{U}\in\mathbf{SU(2)}$
 and introduce the new phases $k'_m$ defined as  $e^{ik_{m}^{\prime}}\left\vert \mathbf{\Phi}_{m}^{\prime}\right\rangle =\hat{U}e^{ik_{m}}\left\vert \mathbf{\Phi}_{m}\right\rangle$  such that the single qubit parametrization remains as   $\left\vert \phi_{m}^{\prime}\right\rangle =\cos\left(\theta_{m}^{\prime}/2\right)\left\vert 0\right\rangle +e^{i\varphi_{m}^{\prime}}\sin\left( \theta_{m}^{\prime}/2\right)  \left\vert 1\right\rangle $.
The new representation for the rotated state $\left\vert \Psi^{\prime}\right\rangle $ is 
\begin{equation} \left\vert \Psi_{(odd)}^{\prime}\right\rangle =A\left(\left\vert \mathbf{\Phi}_{0}^{\prime}\right\rangle +{\textstyle\sum\limits_{m=1}^{\left( N-1\right)/2}}y_{m}e^{ik_{m}^{\prime}-ik_{0}^{\prime}}\left\vert \mathbf{\Phi}_{m}^{\prime}\right\rangle \right).\label{mainU} 
\end{equation} 
In analogy, for an even number of qubits and starting from the state represented in Eq.(\ref{even2}) we arrive to 
\begin{align} \left\vert \Psi_{(even)}^{\prime}\right\rangle  &  =A(\left\vert \mathbf{\Phi}_{0}^{\prime}\right\rangle +y_{1}e^{ik_{1}^{\prime}-ik_{0}^{\prime} }\left\vert \mathbf{\Phi}_{0}^{\prime\bot}\right\rangle \nonumber\\ & +{\textstyle\sum\limits_{\substack{m=2\\(N>2)}}^{N/2}}y_{m}e^{ik_{m}^{\prime}-ik_{0}^{\prime}}\left\vert \mathbf{\Phi}_{m}^{\prime}\right\rangle )\label{evenU}
 \end{align}
  Geometrically speaking, the vectors $\left\{  y_{m}\left\vert \phi_{m}\right\rangle \right\}  $ of the initial state (Eq.(\ref{even2}) or Eq.(\ref{main})) represented on the Bloch ball  simply undergo an uniform rotation under the action of $LU$ operators. In other words the suggested geometric representation of the decomposition rotates as a rigid body. These observations lead naturally to the first criteria offered by our representation: 
\begin{criterion}
If two symmetric states are convertible among each other via $LU$ rotations, their representation on the Bloch ball are identical up to global rotations of the ball.
\end{criterion}
\begin{criterion}
An overcomplete set of $LU$ invariants is formed by the complex numbers $e^{ik_{m}-ik_{n}}\left\langle \mathbf{\Phi}_{n}\right.  \left\vert \mathbf{\Phi}_{m}\right\rangle $, the real positive numbers $\left\{ y_{m}\right\}  $ and the normalization factor $A$.
\end{criterion}
A second option for identifying $LU$ invariants of  state Eq.(\ref{main}) is to apply  $\mathbf{SU(2)}$ operations in order to reduce it into a canonical state characterized by $2N-3$ real parameters \cite{Mandilara}. We define the following $LU-$canonical state for an odd number of qubits \begin{equation} \left\vert \Psi_{\left(odd\right)}^{LU}\right\rangle =A\left(  \left\vert \mathbf{1}\right\rangle +y_{1}\left\vert \mathbf{X}_{1}\right\rangle +{\textstyle\sum\limits_{m=2}^{\left(  N-1\right)  /2}} y_{m}e^{il_{m}}\left\vert \mathbf{X}_{m}\right\rangle \right),\label{SU} \end{equation} and an analogous form for an even number of qubits \begin{equation} \left\vert \Psi_{\left(  even\right)  }^{LU}\right\rangle =A\left(  \left\vert \mathbf{1}\right\rangle +y_{1}\left\vert \mathbf{0}\right\rangle + {\textstyle\sum\limits_{m=2}^{N/2}} y_{m}e^{il_{m}}\left\vert \mathbf{X}_{m}\right\rangle \right).\label{evenSU} \end{equation}
Every state in Eq.(\ref{main}) (Eq.(\ref{even2})) can be reduced to Eq.(\ref{SU}) (Eq.(\ref{evenSU})) by $LU$ transformations. The $2N-3$ real parameters of  Eq.(\ref{SU}) (Eq.(\ref{evenSU})) consists in a complete set of independent $LU$ invariants. 
\begin{criterion}
Two states are equivalent under $LU$ \textit{iff they} have identical $LU-$canonical forms.
\end{criterion}
By definition, states equivalent under $LU$ belong to the same $LU$ class of entanglement and therefore Eqs.(\ref{SU})-(\ref{evenSU}) permits us to identify all the $LU$ classes of generic symmetric states.

Concerning  $IL$ operations the situation naturally becomes more complex. Let us first consider the odd case, and apply a collective operation $\hat{V}\in\mathbf{SL(2,\mathbb{C})}$ to the initial state $\left\vert \Psi_{(odd)}\right\rangle $ in Eq.(\ref{main}). Denoting the resulting state by $\left\vert \Psi_{(odd)}^{^{\prime\prime}}\right\rangle =\hat{V}\left\vert \Psi_{(odd)}\right\rangle$  then \begin{equation} \left\vert \Psi_{(odd)}^{^{\prime\prime}}\right\rangle =A^{\prime\prime}\left(y_{0}^{^{\prime\prime}}e^{ik_{0}^{^{\prime\prime}}}\left\vert \mathbf{\Phi}_{0}^{^{\prime\prime}}\right\rangle + {\textstyle\sum\limits_{m=1}^{\left(N-1\right)/2}}y_{m}^{^{\prime\prime}}e^{ik_{m}^{^{\prime\prime}}}\left\vert\mathbf{\Phi}_{m}^{^{\prime\prime}}\right\rangle\right)  .\label{mainSL}\end{equation} One may observe that \ the action of $IL$ operations preserves the form of the representation, up to a rearrangement of terms which has to be performed, so that the condition $y_{0}>y_{1}>\ldots>y_{\left(N-1\right)/2}$ is satisfied. One can also see that contrary to the $LU$ case, the suggested representation does not provide \textit{obvious} $IL$ invariant quantities. We circumvent this problem by defining the following $IL$-canonical form \cite{Carteret, Mandilara} for a generic state of an odd number of qubits \begin{equation} \left\vert \Psi_{\left(odd\right)  }^{IL}\right\rangle =A\left(  \left\vert\mathbf{0}\right\rangle +\left\vert \mathbf{1}\right\rangle +{\textstyle\sum\limits_{m=2}^{\left(N-1\right)/2}} \lambda_{m}e^{i\xi_{m}}\left\vert \mathbf{\Xi}_{m}\right\rangle \right),\label{SL}\end{equation} where we  applied $\mathbf{SL(2,\mathbb{C})}$ operations to reduce the number of parameters \ in Eq.(\ref{main}) and set : $\left\vert \mathbf{\Phi}_{0}\right\rangle \rightarrow\left\vert \mathbf{0}\right\rangle $, $\left\vert \mathbf{\Phi}_{1}\right\rangle \rightarrow\left\vert \mathbf{1}\right\rangle $ and $y_{1}e^{ik_{1}}\rightarrow1$. We note here that in the general case the coefficients $\lambda_{m}$ in Eq.(\ref{SL}) are not ordered and the condition $\left\vert \lambda_{m}\right\vert <1$ is not necessarily satisfied.

For the case of an even number of qubits $IL$ operations do not conserve the form since they do not preserve the orthogonality condition $\left\langle \phi_{0}\right.  \left\vert \phi_{0}^{\bot}\right\rangle =0$. Despite this fact, we can define an $IL-$canonical form for even number of qubits \begin{align} \left\vert \Psi_{\left(even\right)}^{IL}\right\rangle  &  =\left\vert \Psi_{\left(even\right)}^{IL}\right\rangle =A(\left\vert \mathbf{0} \right\rangle +\left\vert \mathbf{1}\right\rangle +\lambda\left\vert \mathbf{c}\right\rangle \nonumber\\ &  + {\textstyle\sum\limits_{m=3}^{N/2}}  \lambda_{m}e^{i\xi_{m}}\left\vert \mathbf{\Xi}_{m}\right\rangle )\label{evenSL} \end{align} where $\left\vert c\right\rangle =\left(  c\left\vert 0\right\rangle +\left\vert 1\right\rangle \right)  /\sqrt{1+\left\vert c\right\vert ^{2}}$. The complex numbers $c$ and $\lambda$ are not independent and they are related to each other via a parametric relation which is provided in the Appendix together with the proof of Eq.(\ref{evenSL}).

The $2N-6$ real numbers in Eq.(\ref{SL}) (Eq.(\ref{evenSL})) form a complete set of $IL$ invariant and therefore we can  state the following criterion: 
\begin{criterion}
Two states are equivalent under $IL$ operations \textit{iff they } have identical $IL-$canonical forms.
\end{criterion}

By definition, states equivalent under $IL$ operations belong to the same $IL$ class (or just class) of entanglement and these are entangled the same way \cite{Dur}, \cite{Verstraete}. In consequence the canonical forms in Eqs.(\ref{mainSL})-(\ref{evenSL}) permit us to identify all  $IL$ classes for generic symmetric states of qubits as well as representative states of these.

Finally, it is important to note that the decomposition given by Eq.(\ref{ini}) and Eq.(\ref{even}) provides straightforwardly  the Schmidt measure \cite{SN} for every generic symmetric state. Indeed, if we note $r$ the number of non zero $y_m$ coefficients, then the Schmidt measure is given by $P = \log_2(r)$. So, as a byproduct  our method provide for free a method to calculate the Schmidt measure and a classification of entanglement of generic symmetric states according to this widely used measure of entanglement.
\vspace{1cm}
 We illustrate here the different aspects of the  proposed decomposition  by discussing in detail the $3$ qubit case.

According to Eq.(\ref{main}) a generic state of $3$ qubits can be written as  \begin{equation} \left\vert \Psi\right\rangle =A\left(\left\vert \mathbf{\Phi}_{0} \right\rangle +ye^{ik}\left\vert \mathbf{\Phi}_{1}\right\rangle \right) \label{three} \end{equation} where $A$ the normalization factor. The obtained form Eq.(\ref{three}) corresponds to previously derived $3$ qubit extension of Schmidt decomposition \cite{Dur, Acin}. 

\begin{figure}[t]{\centering{\includegraphics*[width=0.3\textwidth]{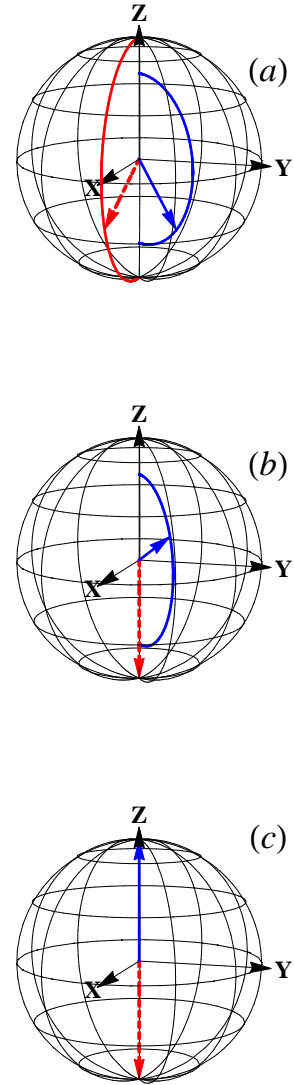}}}\caption{The geometric representation (a) for a general symmetric state of $3$ qubits Eq.(\ref{three}). The normalized vector $\left\vert \phi_{0}\right\rangle $ is represented by the (red) dotted vector while the unnormalized vector  $y\left\vert \phi_{1}\right\rangle $ by a  (blue) solid  vector.   (b) The $LU-$canonic form of  a state as in  Eq.(\ref{threeSU}). (c) The geometric representation of a $GHZ$ state for any number of qubits.} \label{FIG1}\end{figure}

Using our results,  the $3$ qubit symmetric state is represented by the two vectors  $\left\vert\phi_{0}\right\rangle$ and  $y\left\vert \phi_{1}\right\rangle $ in the Bloch ball as in  Fig.1(a). For the  state of maximum tripartite entanglement, i.e. the  $GHZ$ state $\left(\left\vert \mathbf{0}\right\rangle +\left\vert \mathbf{1}\right\rangle \right)  /\sqrt{2}$, we have \ $\left\langle \phi_{0}\right.  \left\vert \phi_{1}\right\rangle =0$ and $y=1$. Therefore,    a $GHZ$ state is geometrically  represented by two normalized and orthogonal vectors (see Fig.1(c)). It is easy to check that this geometric representation  (by two orthogonal vectors)  holds true for all $GHZ$  states independently   of the number ($N$) of qubits. 

Now let us investigate how the 3-tangle $\tau$ \cite{Wootters}, a widely applied measure of entanglement, is related to   the $LU$ invariant characteristics of the representation : $y$,  $e^{-ik}\left\langle \mathbf{\Phi}_{1}\right. \left\vert \mathbf{\Phi}_{0}\right\rangle $ and $A$.  For state $\left\vert\Psi\right\rangle $ in  Eq.(\ref{three}) we have 
\begin{equation} \tau=4y^{2}\left(1-\left\vert \left\langle \mathbf{\Phi}_{0}\right.\left\vert \mathbf{\Phi}_{1}\right\rangle \right\vert ^{2/3}\right)^{3/2}A^{4}.\label{tangle} \end{equation} 

Furthermore, we may compare 3-tangle with the invariant set of parameters deduced by the $LU-$canonical form of the given state $\left\vert \Psi\right\rangle $,i.e., \begin{equation} \left\vert \Psi^{LU}\right\rangle =A\left(\left\vert \mathbf{1}\right\rangle+y\left\vert \mathbf{X}\right\rangle \right)  \label{threeSU} \end{equation} where $\left\vert \mathbf{X}\right\rangle =\left\vert \chi\right\rangle \left\vert \chi\right\rangle \left\vert \chi\right\rangle $ with $\left\vert \chi\right\rangle =\left(\cos\left(\varepsilon/2\right)\left\vert0\right\rangle +e^{i\varphi}\sin\left(\varepsilon/2\right)\left\vert 1\right\rangle \right)$ and 
\begin{equation}A^{-2} =1+y^2+ 2 y \cos (3 \phi ) \sin ^3\left(\frac{\varepsilon}{2}\right).\nonumber\end{equation}
According to Eq.(\ref{SU}) the complete set of $LU$ invariants is formed by the $3$ real numbers $\left\{y,\varepsilon ,\varphi\right\}$ with $0<y\leqslant1$, $0<\varepsilon\leq\pi$ and $0<\varphi\leq2\pi$.  In addition, for this case ($N=3$) the geometric representation for $\left\vert \Psi_{LU}\right\rangle$ in Eq.(\ref{threeSU}) is faithful and one may visualize the set of invariants on the Bloch Ball by the length and position of the vector $y\left\vert \chi\right\rangle $ (see Fig.1 (b)). The $3-$tangle depends on  $\left\{y,\varepsilon,\phi\right\}$ via the simple and intuitive relation \begin{equation}\tau=\frac{4y^{2}\cos^{3}\left(\varepsilon/2\right)}{\left(1+y^2+ 2 y \cos (3 \phi ) \sin ^3\left(\frac{\varepsilon}{2}\right)\right)^2}.\end{equation} The $3$-tangle thus is  monotonically  increasing with $y$ and decreasing with $\varepsilon$, while $\phi$ produces small oscillations. The geometric representation, Fig.1 (b), thus permits us to  compare in a qualitative manner the amount of tripartite entanglement among different states. 
\vspace{1cm}

We have derived a novel representation for generic symmetric states of qubits and we have applied it to provide a complete solution to the problem of entanglement classification  of such states.  We believe that the suggested decomposition is a general tool with potential applications in other fields of quantum physics, including  quantum optics with collective spin states.  

 We thank R. Mosseri and D. Markham for useful discussions. The authors acknowledge financial support by ANR under project HIDE. AM is grateful to Alexander Meill for his critical feedback on the previous version of this work.
\appendix
\section{The decomposition for an odd number of qubits}

Here we prove the existence and the uniqueness of the decomposition \begin{align} \left\vert \Psi_{(odd)}\right\rangle &  =\left({\textstyle\sum\limits_{m=0}^{\left(  N-1\right)  /2}} c_{m}\left\vert \mathbf{\Phi}_{m}\right\rangle \right) \label{init}\\ \left\vert \mathbf{\Phi}_{m}\right\rangle  &  =\left\vert \phi_{m}\right\rangle \otimes\left\vert \phi_{m}\right\rangle ...\left\vert \phi_{m}\right\rangle,\nonumber \end{align} for generic states of an odd number of qubits. The complex amplitudes $c_{m}$ in Eq.(\ref{init}) are of decreasing sequence $\left\vert c_{0}\right\vert>\left\vert c_{1}\right\vert >...>\left\vert c_{\left(  N-1\right)/2}\right\vert $ and $$\left\vert \phi_{m}\right\rangle =\cos\left(\theta _{m}/2\right)  \left\vert 0\right\rangle +e^{i\varphi_{m}}\sin\left( \theta_{m}/2\right)  \left\vert 1\right\rangle $$ single qubit states whicha are either all distinct with each other or all identical.

\begin{proof}

To prove the wanted result, we initially assume  the existence of the decomposition for a given state. Then we derive the conditions under which, the unknown parameters of the decomposition can be uniquely  determined by the Majorana roots of this state.

For  a given state $\left\vert \Psi\right\rangle $ expressed in the computational basis, one can always derive the corresponding Majorana polynomial \cite{Majorana}, $\Psi\left(\alpha\right) =\sum_{m=0}^{N}\lambda_{m}\alpha^{m}$  and the $N$ roots $\left\{\alpha_{n}\right\}$ corresponding to the solution of the equation $\Psi\left(\alpha\right) =0$. In addition, we denote by $\alpha_{0}$ the value of the Majorana polynomial on zero as $\alpha_{0}=\lambda_{0}$.

We assume initially that for the given state  $\left\vert \Psi\right\rangle $, the decomposition,  Eq.(\ref{init}), is possible. For  convenience reasons we rewrite the decomposition Eq.(\ref{init}) as \begin{align} \left\vert \Psi\right\rangle &=\left({\textstyle\sum\limits_{m=0}^{\left(N-1\right)/2}}c_{m}^{\prime}\left\vert \mathbf{\Phi}_{m}^{\prime}\right\rangle \right)\label{ini2}\\ \left\vert \mathbf{\Phi}_{m}^{\prime}\right\rangle  &  =\left\vert \phi_{m}^{\prime}\right\rangle \otimes\left\vert \phi_{m}^{\prime}\right\rangle...\left\vert \phi_{m}^{\prime}\right\rangle\nonumber\end{align} where $\left\vert \phi_{m}^{\prime}\right\rangle$  are now  unormalized vectors : $$\left\vert \phi_{m}^{\prime}\right\rangle =\left(\left\vert 0\right\rangle +\beta_{m}\left\vert 1\right\rangle \right)$$ with $\beta_{m}=e^{i\varphi_{m}}\tan\left( \theta_{m}/2\right)$. The new complex amplitudes $c_{m}^{\prime}$ are related with those in Eq.(\ref{init}) via the relation $c_{m}^{\prime}=c_{m}/\left(1+\left\vert\beta_{m}\right\vert^{2}\right)^{N/2}$.

 At next step, we project the state $\left\vert \Psi\right\rangle $ decomposed as in Eq.(\ref{ini2}) on the spin coherent state $\left\vert \mathbf{\alpha}\right\rangle =\left\vert \alpha\right\rangle\otimes\left\vert \alpha\right\rangle\otimes\ldots\left\vert \alpha\right\rangle$ where $\left\vert \alpha\right\rangle=\left\vert 0\right\rangle+\alpha^*\left\vert 1\right\rangle$. We thus obtain the following polynomial \begin{equation} P\left(  \alpha\right)  ={\textstyle\sum\limits_{m=0}^{\left(N-1\right)/2}}c_{m}^{\prime}\left(1+\alpha\beta_{m}\right)^{N}\label{poly}\end{equation}.

Our aim is to determine the $N+1$ unknown coefficients $\left\{c_{m}^{\prime}\right\}$ and  $\left\{\beta_{m}\right\}$ by the $N+1$ known complex numbers $\left\{\alpha_{n}\right\}$. By construction,  $\Psi\left(\alpha\right)=P\left(\alpha\right)$ since both represent $\left\langle\alpha\right.\left\vert \Psi\right\rangle $. As a direct consequence, the two polynomials share the same roots~$\alpha_i$, such that $P\left(\alpha_{i}\right)=0$ for $i=1,\ldots, N$. This way we obtain the following set of conditions: \begin{equation}{\textstyle\sum\limits_{m=0}^{\left(N-1\right)/2}}c_{m}^{\prime}\left( 1+\alpha_{i}\beta_{m}\right)^{N}=0\label{one}\end{equation} where $i=1,\ldots,N$. In addition,  $\Psi\left(0\right)=P\left(0\right)=\alpha_{0}$ that leads to \begin{equation}{\textstyle\sum\limits_{m=0}^{\left(N-1\right)/2}}c_{m}^{\prime}=\alpha_{0}.\label{two}\end{equation} One may observe that the $N$ conditions Eq.(\ref{one}) form a linear and homogeneous system of equations on the $\left(N+1\right)/2$ parameters $\left\{c_{m}^{\prime}\right\}$. This system is overdefined and in order to have a non-zero solution for $\left\{c_{m}^{\prime}\right\}$  the equations should be linearly dependent.

One way to impose the linear dependence is to require that each vector defined by the $\frac{N+1}{2}$ coefficients $\left[\left(1+\alpha_{i}\beta_{m}\right)^{N}; m=0,1,\cdots \frac{N-1}{2}\right]$ of the $i$th equation, with $i\geq(N+1)/2$, remains in the subspace defined by the first $(N-1)/2$ vectors. This provides us with the following $\left(N+1\right)/2$ conditions: 

\begin{widetext} 

\begin{equation}Det\left[ \begin{array}[c]{cccc}\left(1+\alpha_{1}\beta_{0}\right)^{N} & \left(1+\alpha_{1}\beta_{1}\right)^{N} & \ldots & \left(1+\alpha_{1}\beta_{\left(N-1\right) /2}\right)^{N}\\ \left(1+\alpha_{2}\beta_{0}\right)^{N} & \left(1+\alpha_{2}\beta_{1}\right)^{N} &  & \vdots\\ \vdots & \vdots & \ddots & \vdots \\ \left(1+\alpha_{\left(N-1\right)/2}\beta_{0}\right)^{N} & \left(1+\alpha_{\left(N-1\right)/2}\beta_{1}\right)^{N} & \ldots &\left(1+\alpha_{\left(N-1\right)/2}\beta_{_{\left(N-1\right)/2}}\right)^{N} \\ \left(1+\alpha_{i}\beta_{0}\right)^{N} & \left(1+\alpha_{i}\beta_{1}\right)^{N} & \ldots &\left(1+\alpha_{i}\beta_{_{\left(N-1\right)/2}}\right)^{N} \end{array}\right] =0\label{conDet}\end{equation} 
\end{widetext}
for $i=(N+1)/2,\ldots,N$ which allow the determination of the $\left(N+1\right)/2$ parameters $\left\{\beta_{m}\right\}$. At next step, one may solve the linear system of equations composed by Eq.(\ref{two}) and any $\left(N-1\right)/2$ equation of the system Eq.(\ref{one}), in order to identify the parameters $\left\{c_{m}^{\prime}\right\}$.

One can easily conclude at this point, that the degeneracies in $\beta$'s parameters must be related to those of the roots $\alpha_i$.  Let us postpone the analysis of degeneracies at the end of the proof and show below the uniqueness of the solution assuming that all $\alpha_i$,  (and in consequence all $\beta$'s)  are distinct.

\textsl{Uniqueness of the solution.} What is not obvious from the analysis above is the uniqueness of  the solution   (up to permutations and all $\beta_m$ distinct)   and let us clarify this point.  One can  check that each $i$th determinant Eq.(\ref{conDet}) is of the order $\left(1+N\right)\left(1+3N\right)/8$ on $\beta$'s. However the $i$  determinant  factorizes to a product of ${\textstyle\prod\limits_{\substack{i=0,j=1\\j>i}}^{\frac{N-1}{2}}}\left(\beta_{i}-\beta_{j}\right) $ and a polynomial $f_i=F\left(\left\{\beta_{m}\right\},\left\{\alpha_{n(\neq i)}\right\},\alpha_i \right)$  of order $M=\left(1+N\right)\left(1+3N\right)/8-\frac{\left(\frac{N+1}{2}\right)!}{\left(\frac{N-3}{2}\right)! 2}$. Since a solution where two or more of $\beta$'s coincide is not admissible, one has to  analyze the roots of the polynomials $f_i$. Each of the polynomials $f_i$ is invariant under  permutations of $\left\{\beta_{m}\right\}$ and it can be rewritten on the basis of $\frac{N+1}{2}$ elementary  symmetric polynomials on these variables. In other words, one can proceed with a change of variables from  $\left\{\beta_{m}\right\}$ to symmetric combinations of them. For instance, in the case of $3$ qubits, one rewrites the polynomials in terms of the two variables  $x_0=\beta_0+\beta_1$ and $x_1=\beta_0\beta_1$. Such transformation is further reducing the order of the  polynomials $f_i$ to $K=\frac{N+1}{2}$. According to B\'ezout's theorem, the maximum number of (complex) roots for $K$ polynomials on $K$ variables, of $K$ order each (as we have in our case) is $K^{K}$. However, if one proceeds with the solution of the system of equations one finds only one non trivial solution and the trivial solutions of high degeneracy degree correspond to degeneracies in $\beta$'s. In the simplest case of $3$ qubits the degenerate solution is of third order and implies that $\beta_0=\beta_1$. This fact can be explicitly checked since the polynomials $f_1$ and $f_2$ are  elliptic equations. In the general case ($N>3$), the high degree of degeneracy in the solution is justified by the fact that the polynomials $f_i$ are identical up to an interchange of a single parameter, i.e. the $\alpha_i$ root.

\textsl{Degeneracies in Majorana roots} In the case where one (or more) Majorana root, i.e. $\alpha_k$ is degenerate of $l$th order one has to supplement the system of equations~\eqref{one} and~\eqref{two} with  the following ones \begin{eqnarray}\left.\frac{d^jP(\alpha)}{d\alpha^j}\right|_{\alpha=\alpha_k}& =\nonumber\\{\textstyle\sum\limits_{m=0}^{\left(N-1\right)/2}}c_{m}^{\prime}\beta_{m}^{j}\left(1+\alpha_{k}\beta_{m}\right)^{N-j}&= 0\label{dd}\end{eqnarray} where $j=1,\ldots,l-1$. The solution remains unique in this case and there are no degeneracies in $\beta$'s.

The procedure above does not work when  the highest degeneracy degree ($\gamma$) of the roots $\left\{\alpha_{n}\right\}$ of $\Psi\left(\alpha\right)$ takes an integer value in the interval $\left[\frac{N+1}{2},N-1\right]$. In this case the decomposition Eq.(\ref{init}) is not unique and therefore out of our interest. Let us prove this statement first for the case where $\gamma=\frac{N+1}{2}$ for the root  $\alpha_k$.

The determinant of the system of Eqs.(\ref{dd}) where $j=1,\ldots,(N-1)/2$ supplemented by 

\begin{equation}{\textstyle\sum\limits_{m=0}^{\left(N-1\right)/2}}c_{m}^{\prime}\left( 1+\alpha_{k}\beta_{m}\right)^{N}=0\label{dd2}\end{equation} 

is \begin{equation}{\textstyle\prod\limits_{m=0}^{\frac{N-1}{2}}}\left(1+\alpha_{k}\beta_{m}\right)^{\frac{N+1}{2}}{\textstyle\prod\limits_{\substack{i=0,j=1\\(j>i)}}^{\frac{N-1}{2}}}\left(\beta_{i}-\beta_{j}\right)  =0.\label{dd3}\end{equation} 
This determinant should vanish (to have a non trivial solution for the global system), and therefore either  at least one of the terms $\left(1+\alpha_{k}\beta_{m}\right)$ is zero or at least two of $\beta_m$ coincide. In both cases, the initial system of $\frac{N+1}{2}$ equations Eqs.(\ref{dd})-(\ref{dd2}) contains now $\frac{N-1}{2}$ unknowns. If one pick $\frac{N-1}{2}$ of the equations and  requires that the determinant vanishes, one arrives to an expression similar to Eq.( \ref{dd3}) that implies that  at least one of the terms $\left(1+\alpha_{k}\beta_{m}\right)$ (for $m\neq1$) is zero or at least two of the rest of the $\beta_m$ coincide. Repeating the procedure, $(N-3)/2$ times  one arrives to the conclusion that all $\beta_m$ has to coincide.

Analogous procedure maybe be used to prove that for  $N>\gamma>\frac{N+1}{2}$ the decomposition of the form Eq.(\ref{poly}) does not hold. A straightforward consequence is that the  Dicke states (apart from highest and lowest one) cannot be represented via our proposed decomposition.

It is obvious from the above analysis that the decomposition does not allow degeneracies in $\beta's$ apart from the extreme case (of separable states) where all $\beta's$ coincide. One may prove this result in an  alternative way by searching the conditions under which the decomposition Eq.(\ref{init}) is allowed to have less components. One then arrives to the conclusion, that the existence of a relevant amount of degeneracies in the roots $\alpha$'s is required. However, this analysis leads to a solution that is not unique. Furthermore, we would like to underline   that for some cases, e.g. the $GHZ$ state, less components are present in the decomposition. Such reduced forms are due to vanishing $c$'s and not due to the degeneracies in $\beta's$.

\end{proof}

\section{The decomposition for an even number of qubits }

A generic state of qubits can be put in the following form \begin{align} \left\vert \Psi_{(even)}\right\rangle &  =c_{0}\left\vert \mathbf{\Phi}_{0}\right\rangle +c_{1}\left\vert \mathbf{\Phi}_{0}^{\bot}\right\rangle + {\textstyle\sum\limits_{\substack{m=2\\(N>2)}}^{N/2}}c_{m}\left\vert \mathbf{\Phi}_{m}\right\rangle \label{evenif}\\ \left\vert \mathbf{\Phi}_{m}\right\rangle  &  =\left\vert \phi_{m}\right\rangle \otimes\left\vert \phi_{m}\right\rangle ...\left\vert \phi _{m}\right\rangle \nonumber \end{align} The complex amplitudes $c_{m}$ satisfy now the following conditions: $\left\vert c_{2}\right\vert >...>\left\vert c_{N/2}\right\vert $ and $\left\vert c_{0}\right\vert >\left\vert c_{1}\right\vert $. In addition $\left\langle \phi_{0}\right.  \left\vert \phi_{0}^{\bot}\right\rangle =0$. 

\begin{proof}

Let us assume  that $\left\vert \phi_{0}^{\bot}\right\rangle =\left\vert \phi_{1}\right\rangle $ where $\left\vert \phi_{1}\right\rangle $ a state independent to $\left\vert\phi_{0}\right\rangle$. Then the proof of the existence  of the decomposition Eq.(\ref{evenif}) follows the same lines as for an odd number of qubits. One takes  exactly the same steps  in order to identify the $N+1$  complex numbers $\beta_{m,m\neq1}$ in the polynomial \begin{equation} P\left(  \alpha\right)  ={\textstyle\sum\limits_{m=0}^{N/2}}c_{m}^{\prime}\left(1+\alpha\beta_{m}\right)^{N}\end{equation}
 in terms of the $N$ roots of the Majorana polynomial, $\alpha_0$ and the free parameter $\beta_1$.  At this point one should impose an extra condition and the choice should be such that the final solution is unique up to permutations. The form of dependence of the parameters $\beta_{m,m\neq1}$ on the parameter $\beta_1$ lead us to following condition  \begin{equation} \prod_{i=0}^{N/2}\left( 1+ \beta_i^*\beta_1\right)=0,\label{extra} \end{equation} a condition which notably  remains invariant under the action of local unitary transformations. The solution that follows from the Eq.(\ref{extra}) is not unique since $\left\vert \phi_{1}\right\rangle $ can be orthogonal to any of the rest of states. In other words there are  $N$ different solutions and we have to impose an extra condition making the solution unique. \textit{We choose the solution that  maximizes the amplitude $\left|c_0\right|$ for $\left\vert \mathbf{\Phi}_{0}\right\rangle$.} 

Concerning the impossibility of decomposition Eq.(\ref{evenif}) the same analysis can be made as for the odd number of qubits.  One arrives to the conclusion that when  $N>\gamma\geqslant\frac{N}{2}+1$ one cannot express the state as in Eq.(\ref{evenif}).

\end{proof}

\section{The $IL-$canonical form for an even number of qubits}

A generic state of an even number of qubits can be reduced to the following $IL-$canonical form under the action of invertible local ($IL$) transformations \begin{align} \left\vert \Psi_{\left(even\right)}^{IL}\right\rangle  &  =A(\left\vert\mathbf{0}\right\rangle +\left\vert \mathbf{1}\right\rangle +\lambda\left\vert\mathbf{c}\right\rangle \nonumber\\ &  +{\textstyle\sum\limits_{m=3}^{N/2}}\lambda_{m}e^{i\xi_{m}}\left\vert \mathbf{\Xi}_{m}\right\rangle )\label{evenSL}\end{align} where $\left\vert c\right\rangle =\left(\left\vert 0\right\rangle+c\left\vert 1\right\rangle \right)  /\sqrt{1+\left\vert c\right\vert^{2}}$. The complex numbers $c$ and $\lambda$ are not independent and they are related to each other via a parametric relation.

\begin{proof}

We start by decomposing  the given state as in Eq.(\ref{evenif}). If we divide the expression Eq.(\ref{evenif}) by $c_0$ and renormalize it we arrive to the following equivalent decomposition\begin{align} \left\vert \Psi_{(even)}\right\rangle  &  =A(\left\vert \mathbf{\Phi}_{0}\right\rangle +c_{1}^{\prime}\left\vert \mathbf{\Phi}_{0}^{\bot}\right\rangle +c_{2}^{\prime}\left\vert\mathbf{\Phi}_{2}\right\rangle\nonumber\\& +{\textstyle\sum\limits_{\substack{m=3\\(N>4)}}^{N/2}}c_{m}^{\prime}\left\vert\mathbf{\Phi}_{m}\right\rangle )\label{sl}\\\left\vert\mathbf{\Phi}_{m}\right\rangle &  =\left\vert \phi_{m}\right\rangle \otimes\left\vert \phi_{m}\right\rangle ...\left\vert \phi_{m}\right\rangle \nonumber\\\left\vert \phi_{m}\right\rangle  & =\cos\left(\theta_{m}/2\right)\left\vert 0\right\rangle +e^{i\varphi_{m}}\sin\left(\theta_{m}/2\right)\left\vert 1\right\rangle \nonumber\\\left\vert\phi_{0}^{\bot}\right\rangle  &  =-\sin\left(\theta_{0}/2\right)\left\vert 0\right\rangle +e^{i\varphi_{0}}\cos\left(\theta_{0}/2\right)\left\vert 1\right\rangle \nonumber\end{align} where $A$ the normalization factor.

 An $IL$ transformation is implemented by the action of element of the $SL(2,C)$ group. A matrix representation of such transformation is

\begin{equation}G=\left(\begin{array}[c]{cc}a & b\\d & f\end{array}\right)  \label{ma}\end{equation} where $a$, $b$, $d$, $f$ complex numbers satisfying the condition $af-bd=1$.

We apply transformation $\mathbf{G}=G\times G\times\ldots G$ on the state Eq.(\ref{sl}) and impose that \begin{align}\left(c_{1}^{\prime}\right) ^{1/N}G\left\vert \phi_{0}^{\bot}\right\rangle& =\left(  s\right)  ^{1/N}\left\vert 0\right\rangle \label{c1}\\\left(c_{2}^{\prime}\right)^{1/N}G\left\vert\phi_{2}\right\rangle  & =\left(s\right)^{1/N}\left\vert 1\right\rangle \label{c2}\end{align} where $s$ an arbritary complex number. The conditions Eqs.(\ref{c1})-(\ref{c2}) together with the condition $af-bd=1$ allow us to determine the elements of the matrix $G$ and the parameter $s$ via the known numbers $\theta_{0}$, $\varphi_{0}$, $\theta_{1}$, $\varphi_{1}$, $c_{1}^{\prime}$ and $c_{2}^{\prime}$. 

Under the action of  $\mathbf{G} $ the initial state Eq.(\ref{sl}) is transformed to \begin{align}\mathbf{G}\left\vert \Psi_{(even)}\right\rangle  &  =A^{\prime}(\frac{1}{s}\mathbf{G}\left\vert\mathbf{\Phi}_{0}\right\rangle +\left\vert\mathbf{0}\right\rangle +\left\vert \mathbf{1}\right\rangle \nonumber\\&  +{\textstyle\sum\limits_{\substack{m=3\\(N>4)}}^{N/2}}c_{m}^{^{\prime\prime}}\left\vert \mathbf{\Phi}_{m}^{^{\prime\prime}}\right\rangle ).\end{align}

If we rewrite $\left(\frac{1}{s}\right)^{1/N}G\left\vert \phi_{0}\right\rangle =\lambda^{1/N}\frac{\left(\left\vert 0\right\rangle+c\left\vert 1\right\rangle\right)}{\sqrt{1+\left\vert c\right\vert ^{2}}}$ then it is easy to verify that the complex numbers  $\lambda$ and $c$ are dependent via the following parametric relation \begin{widetext}

\begin{align*}&c=e^{i\phi_{0}}\frac{c_{1}^{\prime}}{c_{2}^{\prime}}\left(e^{i\phi_{0}}\sin\left(\theta_{0}/2\right)\cos\left(\theta_{1}/2\right)-e^{i\phi_{1}}\cos\left(\theta_{0}/2\right)\sin\left(\theta_{1}/2\right)\right)^{-1}\\& \lambda=\frac{\sqrt{1+\left\vert c\right\vert^{2}}}{c_{1}^{\prime}}\frac{\left(e^{i\phi_{0}}\sin\left(\theta_{0}/2\right)\cos\left(\theta_{1}/2\right)-e^{i\phi_{1}}\cos\left(\theta_{0}/2\right)\sin\left(\theta_{1}/2\right)\right)}{\left(e^{i\phi_{0}}\cos\left(\theta_{0}/2\right)\cos\left(\theta_{1}/2\right)+e^{i\phi_{1}}\sin\left(\theta_{0}/2\right)\sin\left(\theta_{1}/2\right)\right)}.\end{align*}

\end{widetext}

\end{proof}

\end{document}